\documentstyle[twoside]{article}
\def\runninghead#1#2{\pagestyle{myheadings}
\markboth{\hfill{\protect\footnotesize\it{\quad #1}}}
{{\protect\footnotesize\it{#2\quad}}\hfill}}
\begin{document}

\newcommand{\Vi}{\left(\sum\limits_i{\bf V}_i\cdot\nabla\right)}

%FOR CREATING THE OPENING PAGE NUMBER
\def\fpage#1{\begingroup
\voffset=.3in
\thispagestyle{empty}\begin{table}[b]\centerline{\footnotesize #1}
        \end{table}\endgroup}

\thispagestyle{empty}\setcounter{page}{1}
\vspace*{0.88truein}
\fpage{1}

\runninghead{\bf A. E. Chubykalo}
{\bf Comment on "Do Zero-Energy ..."}

$$$$
{\large {\bf COMMENT ON ``DO ZERO-ENERGY SOLUIIONS OF MAXWELL
EQUATIONS HAVE THE PHYSICAL ORIGIN SUGGESTED BY A. E. CHUBYKALO?" BY V.V.
DVOEGLAZOV}
\bigskip
\bigskip

$\;\;\;\;\;\;\;\;\;\;${\bf A. E. Chubykalo}

\bigskip

$\;\;\;\;\;\;\;\;\;\;${\it Escuela de F\'{\i}sica, Universidad Aut\'onoma
de Zacatecas

$\;\;\;\;\;\;\;\;\;\;$Apartado Postal C-580\, Zacatecas 98068, ZAC., M\'exico}

\bigskip
\bigskip

$\;\;\;\;\;\;\;\;\;\;$Received October 6, 1999

\baselineskip 5mm

\bigskip

$$$$

\bigskip


\bigskip
\noindent
Dvoeglazov's paper is not trivial because  we know the
classical definition of any non-trivial truth: a truth is not trivial {\it
if and only if} an {\it opposite} statement is not {\it false} (and vice
versa, of course).  So I strongly recommend his paper for a reader.

But I have a few notes:

1) V. Dvoeglazov writes in his footnote number 7: "In my opinion, equations
(20,21) of ref.~\cite{Chub} are just another form of the Maxwell
equations for this particular case, in the sense that there is no new
physical content if one expects that the Maxwell electrodynamics describes
also the Coulomb interaction."

The point is that precisely this (``{\it another}") form of the Maxwell
equations (namely with {\it total} derivatives!) allows us to describe
the Coulomb interaction as an {\it action-at-a-distance} (it was shown in
\cite{Chubo}).  It is well-known that {\it generally accepted} Maxwell
equations (with {\it partial} derivatives only) describe just
transverse\footnote{{\large ``transverse" in this case means that vectors
{\bf E} and {\bf B} are perpendicular with respect to wave vector {\bf k}
in every point.}} electromagnetic waves in vacuum which spread with a
limited velocity (i.e. they describe so called {\it short-range}
interaction only).  That is why there is {\it new physical
content}{\footnote{\large I should to note here that a considerable number
of papers have recently been published which declare (prove?) an existence
of so called longitudinal electromagnetic waves in vacuum but these ideas
still are not generally accepted in classical electrodynamics (see, e.g.,
review \cite{Dvo}).}} in Eqs.  (18-21) \cite{Chub}.  Note also that V.
Dvoeglazov here erroneously mentioned just Eqs.  (20,21).  The point is
that Eqs.  (20,21) describe exclusively instantaneous
(action-at-a-distance) interaction while the whole system of our equations
(18-21) (Eqs. (1-4) in this Comment) describe both instantaneous and
short-range.  These equations are written as two {\it uncoupled} pairs of
differential equations:

\begin{eqnarray} && \nabla\times{\bf
H}^*=\frac{1}{c}\frac{\partial{\bf
E}^*}{\partial t},\\ && \nabla\times{\bf
E}^*=-\frac{1}{c}\frac{\partial{\bf B}^*}{\partial t}
\end{eqnarray}
and
\begin{eqnarray} && \nabla\times{\bf
H}_0=\frac{4\pi}{c}{\bf j} -\frac{1}{c}\Vi{\bf E}_0,\\ && \nabla\times{\bf
E}_0=\frac{1}{c}\Vi{\bf H}_0.
\end{eqnarray}
This system follows from:

\begin{eqnarray} && \nabla\times{\bf H}=\frac{4\pi}{c}{\bf j}+\frac{1}{c}
\frac{d{\bf E}}{dt}\\ && \nabla\times{\bf E}=-\frac{1}{c}\frac{d{\bf
H}}{dt} \end{eqnarray}
where
\begin{eqnarray} && {\bf E}={\bf E}_0+{\bf E}^*={\bf E}_0({\bf
r}-{\bf r}_q(t))+ {\bf E}^*({\bf r},t),\\ && {\bf H}={\bf H}_0+{\bf
H}^*={\bf H}_0({\bf r}-{\bf r}_q(t))+ {\bf H}^*({\bf r},t),
\end{eqnarray}
{\bf r} is the fixed point of observation, ${\bf r}_q$ is the point of the
location of a moving charge $q$ at the instant $t$, the total time
derivative of any vector field value {\bf E} (or {\bf H}) can be
calculated by the following rule:  \begin{equation} \frac{d{\bf
E}}{dt}=\frac{\partial{\bf E}^*}{\partial t}-\Vi{\bf E}_0, \end{equation}
here ${\bf V}_i$ are
velocities of the particles at the same moment of time of
observation.\footnote{{\large Note (see \cite{Chubo}) that unlike the
fields $\{\}^*$ the fields $\{\}_0$ {\it do not retard.}}}

2) V. Dvoeglazov's phrase: "The main problem with the
Chuby-kalo derivation is the following:  the integrals are {\it
divergent} when they extend over all the space. ... But, to the best of my
knowledge, some persons claimed that such procedures do not have sound
mathematical foundation."

I do not see any {\it mathematical} prohibition to use such procedures.
See, e.g., \cite{Fih} (Vol. 2, ch. 17, \S 1, point 282).  Fihtengoltz here
considers the following {\it definition} of the integral with infinite
limits:  \begin{equation} \int\limits_a^\infty
f(x)dx=\lim\limits_{A\rightarrow\infty} \int\limits_a^A f(x)dx.
\end{equation}
If this limit ({\it rhs} 10)  does not exist {\it or}(attention!) is
infinite, they say that this integral {\it diverges}. But then Fihtengoltz
adds, however, that in the case when the limit ({\it rhs} 10) {\it exists}
and it is {\it equal} to infinity {\it one can consider} the infinite
limit ({\it rhs} 10) as a {\it value} of the integral ({\it lhs} 10).

V. Dvoeglazov asks: "Furthermore, even if one accepts its validity
the total energy resulting from integration of (28) over the whole space is
to be infinite!?"\footnote{{\large Recall that Eq. (28) in \cite{Chub} is:
$$
w=\frac{2{\bf E}^*\cdot{\bf E}_0+2{\bf
H}^*\cdot{\bf H}_0+ E^2_0+H^2_0}{8\pi}.
$$}}

Why not? Many years ago, in Newton's times, for example, nobody (I mean,
physicists) doubted that the Universe had an infinite number of stars. It
meant that the mass of the Universe was infinite. And {\it this} mass is
no more than $\int\limits_\infty \varrho d{\cal V}$ over the infinite
volume, where $\varrho$ is a {\it limited} mass density of the Universe.
{\it Now} many physicists {\it believe} that the mass (and, of course, the
total energy) of our Universe is limited but their statements have no any
evidence (however the statement that the Universe has an infinite mass
{\it also} has no evidence). Unfortunately, it is still a matter of
the belief.

At last V. Dvoeglazov cites: "It was noted by Barut~\cite[p.105]{Barut}
that in the case of non-vanishing fields at the spatial infinity `{\sl we
cannot expect to find globally conserved quantities}'."

I have only one question here: Is it possible {\it experimentally} to find
{\it globally} conserved quantities?

3) The final phrase of V. Dvoeglazov's paper is: "... we are not yet
convinced in the necessity  of correction of the formula for the energy
density for radiation field because of the absence of firm experimental
and mathematical bases in \cite{Chub}."

In turn, I explained in this short comment that if we {\it suppose} that
the radiation field {\it exists} in infinity ({\it mathematically} it
is possible, see above, on the other hand the opposite statement has no
evidences) we must correct the formula of the energy density for the
electromagnetic field.

$$$$ {\large {\bf REFERENCES}}

}

\end{document}